# A Robust Quantitative Comparison Criterion of Two Signals based on the Sobolev Norm of Their Difference

Marc Perlin[1,*] and Miguel D Bustamante[2,†]

[1]Naval Architecture & Marine Engineering, University of Michigan, Ann Arbor, MI 48108 USA

[2] Complex and Adaptive Systems Laboratory, School of Mathematical Sciences, University College Dublin, Belfield, Dublin 4, Ireland

22 December 2014

## Abstract

We present a method for the quantitative comparison of two surfaces, applicable to temporal and/or spatial extent in one or two dimensions. Often surface comparisons are simply overlaid graphs of results from different methodologies that are qualitative at best; it is the purpose of this work to facilitate quantitative evaluation. The surfaces can be analytical, numerical, and/or experimental, and the result returned by the method, termed surface similarity parameter or normalized error, has been normalized so that its value lies between zero and one. When the parameter has a value of zero, the surfaces are in perfect agreement, whereas a value of one is indicative of perfect disagreement. To provide insight regarding the magnitude of the parameter, several canonical cases are presented, followed by results from breaking water wave experimental measurements with numerical simulations, and by a comparison of a prescribed, periodic, square-wave surface profile and the subsequent manufactured surface.

---
[*] E-mail: perlin@umich.edu (corresponding author)
[†] E-mail: miguel.bustamante@ucd.ie

**Keywords: Quantitative Surfaces Comparison, Sobolev Norm, Surface Similarity Parameter, Normalized Error, Canonical Surfaces Comparison**

## I. Introduction

This manuscript demonstrates a method of **quantitative** comparison of two surfaces (temporal or spatial). It is stressed that the method presented is generic and can be implemented to compare surfaces in three dimensions (i.e. two dimensional/2D surfaces) as well as two dimensions (i.e. one dimensional/1D surfaces) and that the data can be spatial as well as temporal. Herein results are presented only for 1D surfaces.

In the vast majority of high-quality research articles, the comparison of signals (such as experimental versus numerical versus theoretical data points) is done by simply overlaying them. This is of course a purely qualitative comparison, which in many cases gives more than enough evidence to support the underlying analytical treatment of the problem. However, it is possible to provide more quantitative comparisons of two or more signals, and it is desirable to do so in cases when there are competing theories or numerical methods or experimental setups. Historically, the comparison of signals determined the invention of modern statistical concepts such as regression and correlation [1, 2]. These methods form part of the general set of statistical methods such as comparison of mean, standard deviation and high-order moments, skewness, kurtosis, amplitude spectra, etc. [3, 4, 5]. Statistical methods are very useful when the data set is strongly influenced by a probability distribution (typically normal), which is the case in the medical sciences for example, or in the case of surface measurement when the techniques have significant experimental error [6]. However, in the present discussion we will consider the situation when the experimental signal's error is not significant and therefore the discrepancy between this signal and a numerically- (or theoretically-) generated signal is mainly due to the limitation of the numerical method, experimental technique, and/or the analytical model used. What we propose in this paper is a direct comparison between the signals, without assuming any hypotheses on the signals' probability distributions. In this sense our proposed method is in the spirit of Nicho [7].

Notice that comparison methods such as correlation between two functions are too general and do not use the fact that we are comparing two signals that can be subtracted. In other words, comparators such as correlation allow us to compare "apples with oranges"; in particular, subtraction between two functions is not used or even defined in the computation of a correlation. On the other hand, a sensible test should distinguish clearly the case "$f(t)$ versus $f(t)$" (exact correlation, equal functions) from the case "$f(t)$ versus $-f(t)$" (perfect anti-correlation and unequal functions). Our proposed definition via a dimensionless quotient $Q \in [0,1]$ does distinguish the two cases, giving $Q = 0$ for the first case and $Q = 1$ for the second case. However it should be stressed that our quotient $Q$ is computed in terms of the difference between two quantities, which cannot be reduced to a correlation.

Unfortunately, quantitative comparisons of time (or space, i.e. surface) series, although available, are computed rarely. Herein we use the Sobolev norm, $H^s$, a particular case of which is the $L^2$-norm, obtained when $s = 0$ so the weighting function that includes frequency or wavenumber has been set to unity. Most importantly, this distance norm includes phase as well as amplitude. This is accomplished via Fourier transform which obtains a complex value for each frequency/wavenumber in each of the time series, and then takes the modulus of their difference. The resulting value, once normalized appropriately (e.g. using the triangle inequality) provides a value between zero and one, with zero representing perfect agreement and one representing completely dissimilar signals.

## II.  **Methodology**

For functions of one variable, norms defined in the Sobolev space $W^{k,p}$ with $p = 2$ are used in conjunction with a normalizing denominator via the triangle inequality. To begin, the Sobolev norm $\|f(\cdot)\|$ of a function $f(t)$ is defined by Evans [8] as

$$\|f(\cdot)\|^2 = \int_{-\infty}^{\infty} |F(\omega)|^2 (1 + |\omega|^2)^k d\omega$$

where $F(\omega)$ is the usual Fourier transform of $f(t)$, the term associated with the parentheses in the integrand is a weighting factor that depends on $k$ and is used typically to weigh more heavily the higher frequencies. Here, as there is no apparent reason to weigh one frequency more than another, the entire term is taken to be unity by choosing $k = 0$. Of course, " $\cdot$ " and $\omega$ are generic, but herein usually represent time and radian frequency or space and wavenumber, respectively.

### A.  Error between two signals

Usually in using the $L^2$ or Sobolev norms for comparison between two signals $f(t)$ and $g(t)$, one takes the difference between the squared norms, which in the above case with the assumptions given would yield

$$(Error_0)^2 = \|f(\cdot)\|^2 - \|g(\cdot)\|^2 = \int |F(\omega)|^2 \, d\omega - \int |G(\omega)|^2 \, d\omega. \qquad (1)$$

While this is useful when the two signals have been obtained through different methods so that they cannot be compared pointwise in time or frequency, it is clear that a more quantitative comparison between the two signals would be provided by the following error estimate:

$$(Error_1)^2 = \|f_1(\cdot) - f_2(\cdot)\|^2 = \int |F_1(\omega) - F_2(\omega)|^2 \, d\omega. \qquad (2)$$

Note that this definition has the very important advantage of using the phase information as well as the amplitude in the frequency domain to evaluate and quantify the difference between two 1D surfaces, and can easily be extended to the difference between two 2D surfaces.

B.  Normalized Error: Surface Similarity Parameter

Notice that the error definition, (2) is widely used for fitting functions via the least-squares approximation. However, to be useful as a surface comparator, one needs a quantity that does not change if the signals get re-scaled by a common constant factor. In other words, one needs to define a relative error, devoid of physical dimensions. We achieve this by using the triangle inequality to normalize the error, and define a quotient that varies between zero (for perfect agreement) and one (for perfect disagreement):

$$Q(f_1, f_2) \equiv \frac{\|f_1(\cdot) - f_2(\cdot)\|}{\|f_1(\cdot)\| + \|f_2(\cdot)\|}, \qquad (3)$$

or explicitly,

$$Q(f_1, f_2) = \frac{\left(\int |F_1(\omega) - F_2(\omega)|^2 \, d\omega\right)^{1/2}}{\left(\int |F_1(\omega)|^2 \, d\omega\right)^{1/2} + \left(\int |F_2(\omega)|^2 \, d\omega\right)^{1/2}}.$$

We call the above quotient "normalized error" or "surface similarity parameter". This is similar in spirit to the "normalized Euclidean distance", defined as $E(f_1, f_2) \equiv \frac{\|f_1(\cdot) - f_2(\cdot)\|}{\sqrt{2}\sqrt{\|f_1(\cdot)\|^2 + \|f_2(\cdot)\|^2}}$ and implemented in commercial packages such as *Mathematica*[‡]. Both quotients $E$ and $Q$ are between 0 and 1. They satisfy the relation

$$\frac{E(f_1, f_2)^2}{Q(f_1, f_2)^2} = \frac{1}{2} + \frac{\|f_1(\cdot)\|\|f_2(\cdot)\|}{\|f_1(\cdot)\|^2 + \|f_2(\cdot)\|^2} . \quad (4)$$

We choose to use $Q(f_1, f_2)$ because of its simpler presentation and also because it penalizes more when the functions $f_1, f_2$ disagree, due to the inequalities

$$\left(\frac{Q(f_1, f_2)}{\sqrt{2}} \leq\right) E(f_1, f_2) \leq Q(f_1, f_2) .$$

This can be derived from (4). In the next section the model is exercised to give the reader a sense of the expected results when using the surface similarity parameter, (3). Some results from the other error definitions, (1), (2) are presented also.

C.  Resolution dependence of surface similarity parameter $Q(f_1, f_2)$

In practical applications data are available at discrete points only. Consequently the normalized error, (3) has to be implemented in a discrete sense. This carries an inherent error [i.e., error bars on the evaluation of $Q(f_1, f_2)$] which depends on the number of data points. Actually this is a good feature as the error bars provide more reliability to a comparison.

All this can be shown for the simple example of a square signal representation. Consider the square signal defined on the domain $[0, 2\pi) \ni t$:

---

[‡] http://reference.wolfram.com/language/ref/NormalizedSquaredEuclideanDistance.html

$$f(t) = \begin{cases} 1 & if\ t \in [0, \pi), \\ -1 & if\ t \in [\pi, 2\pi). \end{cases}$$

The Fourier representation for this signal, truncated to Fourier mode $(2p + 1)$, is

$$f_p(t) = \frac{4}{\pi} \sum_{n=0}^{p} \frac{1}{2n+1} \sin(2n+1)t.$$

It follows $\lim_{p\to\infty} \|f_p(\cdot) - f(\cdot)\| = 0$, where we use for simplicity the $L^2$-norm in the spatial representation:

$$\|f(\cdot)\|^2 \equiv \int_0^{2\pi} f(t)^2 dt.$$

Let us consider the normalized error between the truncated sum $f_p(t)$ and the squared signal. As preliminary results, we obtain readily the following formulae:

$$\|f_p(\cdot) - f(\cdot)\|^2 = \frac{16}{\pi} \sum_{n=p+1}^{\infty} \frac{1}{(2n+1)^2} = \frac{4}{\pi} \frac{d^2}{dp^2} \log\left[\Gamma\left(p + \frac{3}{2}\right)\right],$$

$$\|f_p(\cdot)\|^2 = \frac{16}{\pi} \sum_{n=0}^{p} \frac{1}{(2n+1)^2} = 2\pi - \|f_p(\cdot) - f(\cdot)\|^2,$$

$$\|f(\cdot)\|^2 = 2\pi,$$

where $\Gamma(z)$ is the Euler gamma function. In the limit as $p$ becomes very large we have the following approximation:

$$\|f_p(\cdot) - f(\cdot)\|^2 \approx \frac{4}{\pi(p+1)} + O\left(\frac{1}{(p+1)^3}\right).$$

Therefore, the surface similarity parameter becomes, in that limit,

$$Q(f_p, f) \approx \frac{1}{\pi\sqrt{2p}} + O\left(\frac{1}{p^{\frac{3}{2}}}\right).$$

The same asymptotic behavior is observed for the normalized Euclidean distance $E(f_p, f)$.

The square signal example shows that the more terms we take, the closer the two functions $f_p, f$ are, and correspondingly the quotient $Q(f_p, f)$ tends to zero. This example compares an "exact" function $f$ with an "approximation" $f_p$ with limited resolution $(2p + 1)$. Therefore, $Q(f_p, f)$ is a good measure of the error that we incur, which depends on the maximum frequency resolved. We interpret this error as "inherent" (i.e. leading to error bars). For example, when $p = 2000$ (corresponding to 4001 modes resolved) we obtain $Q(f_p, f) \approx \frac{1}{\pi\sqrt{4000}} \approx 5 \times 10^{-3}$.

This exercise leads to the following definition of error bars for a general comparison between two functions using a limited resolution $N$:

$$[\text{Error bars on } (f_1, f_2)] = \pm \frac{1}{\sqrt{N}}.$$

## III. Exercising the Model: Canonical and Physical Comparisons

To demonstrate the surface similarity parameter/normalized error $Q(f_1, f_2)$ several canonical cases are examined. First, the simplest cases of two time series are presented, each with the same amplitude but different phase – in-phase or $\pi$ radians out-of-phase. Second, cases where the amplitudes vary but the phases are the same, and vice versa, where the phases vary but the amplitudes are the same are given. Third, we show canonical results from two time series with random amplitudes and phases, and with the same amplitudes and random phases only. Subsequently, in this section for comparison of signals, we use actual experimental and numerical data from Tian, Perlin, and Choi [9] and from Tian and Choi [10]. The first of these papers investigated dispersive focusing of steep and breaking waves experimentally and numerically using an eddy-viscosity model; the latter manuscript included the addition of wind forcing. To compare the results, as is common, the manuscripts plotted the numerical simulations and the physical measurements on the same axes. In general the results of the simulations compared "well" with the experiments; however, a method to compare the results quantitatively was not presented. Lastly, we compare a prescribed, structured surface (rectangular/square waves with different duty cycles manufactured for use in flow tunnels to investigate roughness effects) to its manufactured counterpart. The measurement of the manufactured surface was made using a LEXT laser interferometer. The manufacturing process was that of silicon wafers.

A. Comparisons of Sinusoids

As a first performance test on the norm, two sine waves of the same amplitude and frequency that were in-phase and that were $180^\circ$ out-of-phase were examined. As expected, as these two examples are perfectly in-phase and out-of-phase, the former gave a surface similarity parameter $Q$ of zero, while the latter generated a value of one. The same values result when additional sine waves are added for each of the two signals with the same set of phases. Obviously this is also the case for two or more sine waves of the same frequency, but different amplitudes, when they are in-phase and $180^\circ$ out-of-phase. And, also as expected, one can generate any value between zero and one with appropriately chosen amplitudes and phases. For several waves with varying frequencies and amplitudes, the quantitative comparison is more interesting as shown below.

B. Comparisons of Signals with Random Phase Shifts of Vastly Differing Magnitudes

To assist with intuition as regards the magnitude of the surface similarity parameter, $Q$, time series with 50 frequency components in the spectrum were determined. And to resemble somewhat an ocean spectrum, the frequencies were divided into bins of 10 components, each bin with pseudo-random amplitudes, but decreasing multiplier. That is, the lowest frequency bin of 10 components had a multiplier of 20, the next higher frequency bin amplitudes were multiplied by 10, then by seven, then by three, and finally by one. The mean/DC component was set to zero. In the second time series, the spectral amplitudes were identical with the first. However, the phase shifts for the first time series were chosen pseudo-randomly between zero and $2\pi$, while the phase shifts for the second time series were those of the first time series plus a perturbation parameter, $\varepsilon$. The perturbation parameter was set to 0.01, 0.10, 0.50, and 1.0, respectively, and the surface similarity parameter, $Q$, was computed. As expected, as $\varepsilon$ approaches one, $Q$ approaches one. The table below provides the results of the comparison. Obviously, for different pseudo-random-number-generator seed values, these numbers would vary somewhat, but would remain the same order of magnitude.

| $\varepsilon$ | 0.01 | 0.10 | 0.50 | 1.0 |
|---|---|---|---|---|
| $Q$ | 0.0169 | 0.1674 | 0.6670 | 0.7145 |

**Table**. The four values of the perturbation, $\varepsilon$, used to generate signals with vastly different random phase shifts and the resulting normalized error, $Q$.

To show graphically a comparison of two sets of the time series, we present two of the cases: $\varepsilon = 0.10$ and 1.0 in Figure 1.

C. Comparison of the Numerical and Experimental Results of Tian et al. [9] and of Tian and Choi [10]

Reproduced below in Figure 2 are three graphs from Tian et al., part of their Figure 13, case W3G4. This figure compares the experimental results, which include a breaking wave due to wave focusing, with the numerical simulations of the same focusing where an eddy viscosity was implemented to remove energy from the wave field at an appropriate time. As published, the results were said to agree well, and as can be seen this is the case. Here, we use the normalized error to make a quantitative comparison. A pre-breaking time series comparison ($x = 12.59$m) is shown in the upper inset, followed by two cases subsequent to breaking ($x = 14.38$m and $x = 17.20$m). First, as is evident from reviewing the figures of time series measured at the two downstream locations subsequent to breaking, it is seen that the relative difference between the experiments and the simulations improve further downstream of the break point. It is noted also that the $x = 12.59$m time series from the experiment and from the simulation exhibit the "steepest-in-time" sea-surface elevation, and show phase as well as amplitude differences.

These surfaces were used as input to the surface similarity parameter/normalized error calculated in Matlab using (3) and the subroutine "norm" appropriately, and both yielded the following results. For $x = 12.59$m, the surface similarity parameter was found to be $Q = 0.179$; for $x = 14.38$m, the surface similarity parameter was $Q = 0.158$; for $x = 17.2$m, the surface similarity parameter was determined to be $Q = 0.131$. Recall that a value of zero indicates perfect agreement, i.e. the same signal. These quantitative results are in agreement with the qualitative observations mentioned previously.

In a second set of experiments, Tian and Choi [10] reported wind effects on breaking waves; from their data we extract a breaking case without wind. Their Figure 7b is reproduced here as Figure 3, shown below with four positions: 2.84, 5.13, 7.04, and 9.07m. As is evident qualitatively from the figures, the difference between the two time-series' signals increases in the downstream direction. Using the surface similarity parameter/normalized error $Q$ as described above yields values of 0.075, 0.149, 0.174, and 0.188, respectively, for the downstream measurement locations. Clearly, these quantitative results are in agreement with our general evaluations by "eye", but provide a substantiation and quantification of them.

Of course as expected, the larger the phase shift difference, the less the two time series resemble each other, and the larger the $Q$ value.

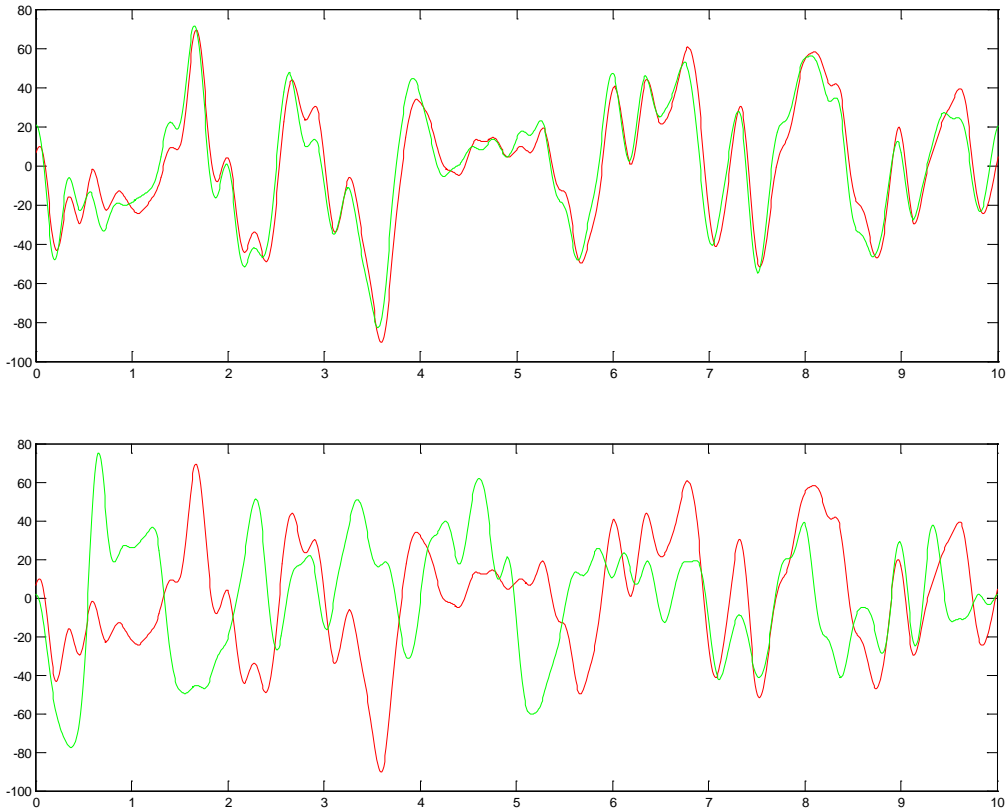

**Figure 1**. A comparison of two of the four sets of time series with random phase shifts. Presented are perturbed cases $\varepsilon = 0.10$ and $1.0$.

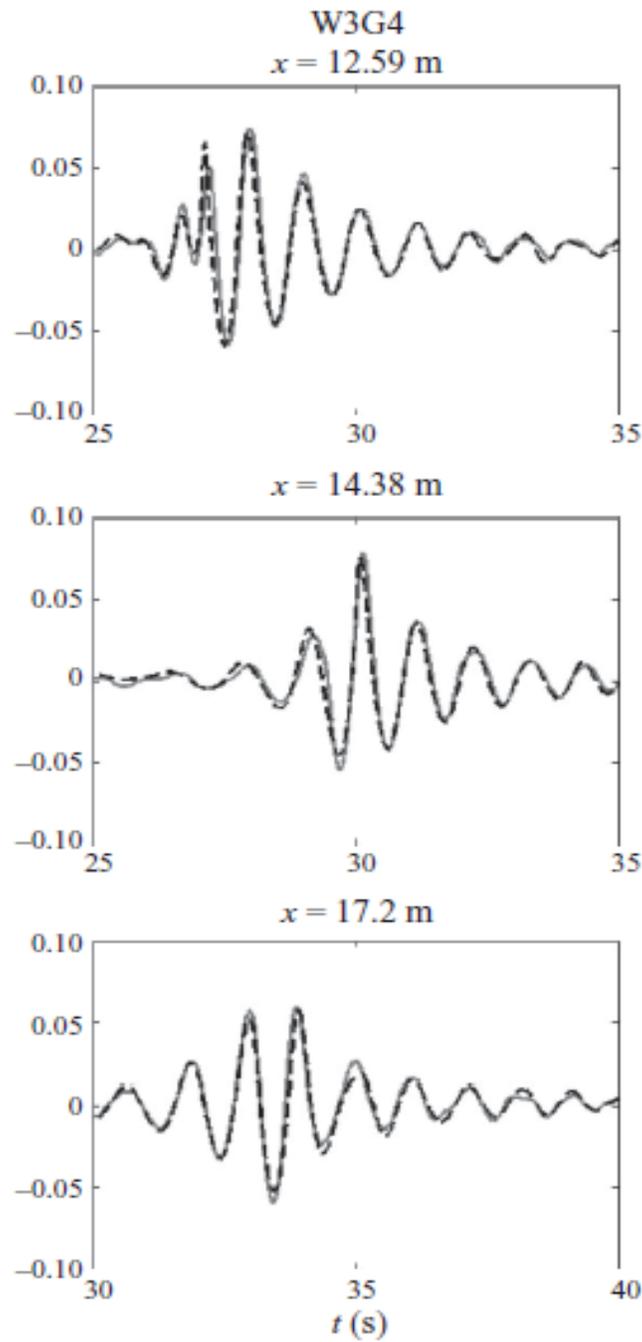

**Figure 2**. Results from Tian et al. [9], part of their Figure 13, is shown. The upper inset is results from pre-breaking measurements recorded 12.59m downstream. The middle and lower insets are post-breaking measurements recorded 14.38 and 17.20 meters downstream of the wave maker. The numerical simulations are represented by the dashed curve, while the experimental measurements are shown by the solid curve. An eddy viscosity model was used to model breaking.

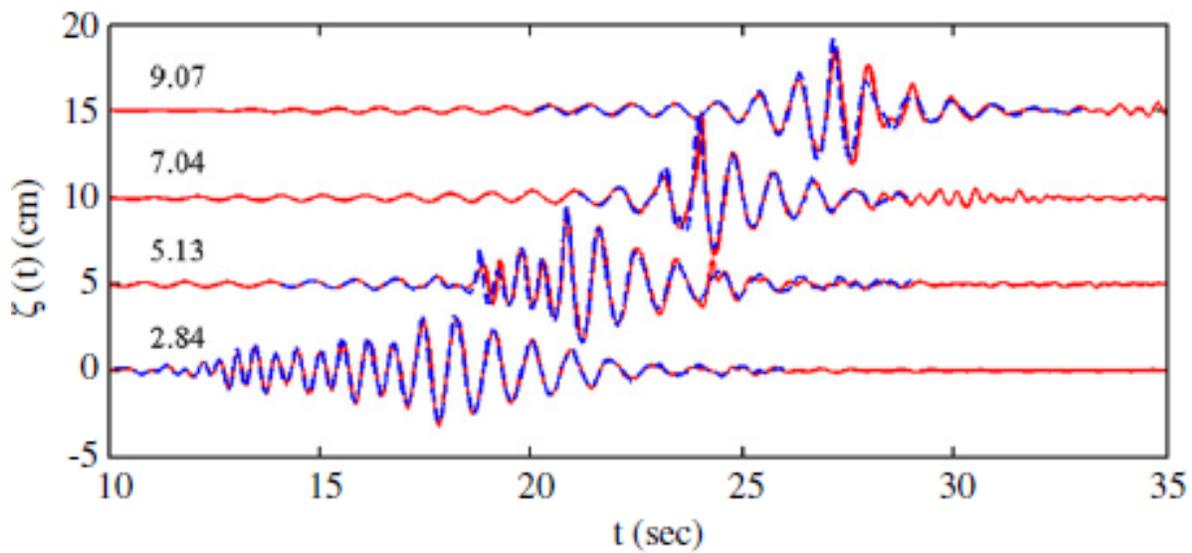

**Figure 3**. Figure 7b from Tian and Choi [10] is reproduced here. Presented are four positions, 2.84, 5.13, 7.04, and 9.07m, respectively, from the wave generator, and in the absence of wind forcing. The experimental data are shown as solid curves while the numerical simulations are represented by the dashed curves.

D.   Comparison of Prescribed and Resulting Square-Wave Manufactured Surfaces

As a final comparison of surfaces, we present a prescribed periodic, spatial surface for a surface over which flow will be measured to determine roughness effects on inner boundary layers, and that of the resulting manufactured surface as measured using a laser interferometer. The laser interferometer used is a LEXT device with resolution on the order of a nanometer. In Figure 4 we present a graph of the prescribed surface: the red curve, which a square wave with a wavelength of 1000μm, a peak-to-peak height of 100μm and a trough length of 900μm (i.e. a square wave with a duty cycle of 10%). Superposed on this same figure in green is the measured surface of a manufactured specimen (as done with silicon wafers) meant to match the specified, red feature. As can be seen in the figure, the two surfaces are similar as expected. But how quantitatively similar or dissimilar are the data? Using the surface similarity parameter/normalized error $Q$ on these data produces a value of 0.191, which once again quantifies their relationship.

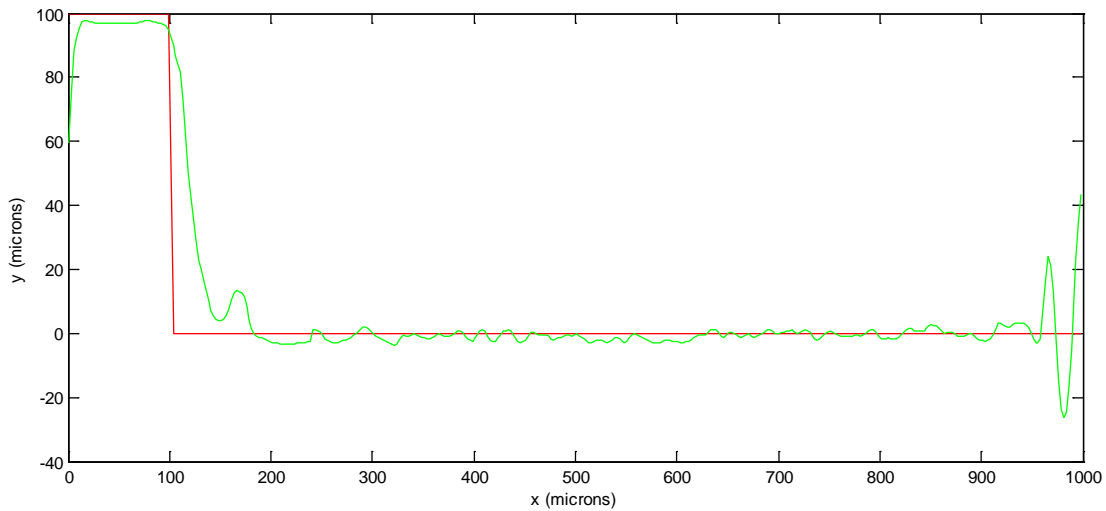

**Figure 4**. A prescribed rectangular-wave surface, and its manufactured counterpart. The manufactured, silicon wafer surface has been measured using a LEXT interferometric instrument.

## IV. Conclusions

We have presented a quantitative method to compare temporal or spatial series in one dimension, or temporal or spatial surfaces in two dimensions, by introducing a "surface similarity parameter", which is a normalized error between two surfaces written in terms of Sobolev norms. Similar to some extent with other coefficients used for this purpose in that the magnitude of the surface similarity parameter varies between zero and one, but very different in one important way: this parameter includes contributions due to both the amplitude and the phase differences.

In the second section the surface similarity parameter was described in detail, and its sensitivity to the discreteness of the compared data was discussed. Following this, the method was exercised using canonical series (sinusoidal profiles with various phase shifts) to give the reader a sense for the magnitude of the parameter versus usual comparison techniques (e.g. graphing) and their interpretation. Then, the method was used with numerical and actual experimental data of breaking water waves, and the results were seen to be consistent with the qualitative ones, however they were quantitative in nature. Last, prescribed surface structures of rectangular waves with varying duty cycles and actual manufactured surfaces intended to duplicate the prescribed surfaces (that were measured with great accuracy using an interferometry system) were compared using the surface similarity parameter. Again the reader was given a sense of the parameter as a function of the discrepancies between prescribed and actual surfaces.

The surface similarity parameter provides a quantitative measure of the discrepancies between a set of 1D surfaces as well as between a set of 2D surfaces. The surfaces considered in the comparison can be experimental, numerical, or analytical, and can be any combination of one or two types thereof. And importantly, the method generates numerical values that are interpreted easily as they lie between zero and one.

## Acknowledgements

MP acknowledges sabbatical leave granted by the University of Michigan, and thanks the University College Dublin for providing an office and associated necessities. MDB acknowledges financial support from Science Foundation Ireland (SFI) under Grant No. 12/IP/1491.